\documentclass[twocolumn,showpacs,preprintnumbers,amsmath,amssymb]{revtex4-1}

\usepackage{graphicx}
\usepackage{amsmath}
\usepackage{amssymb}
\usepackage{bm}
\renewcommand{\vec}[1]{\boldsymbol{#1} }

\begin{document}


\title{Pattern formation in wet granular matter under vertical vibrations}

\author{Lorenz Butzhammer}
\author{Simeon V\"olkel}
\author{Ingo Rehberg}
\author{Kai Huang}
\email{kai.huang@uni-bayreuth.de}
\affiliation{Experimentalphysik V, Universit\"at Bayreuth, 95440 Bayreuth, Germany}

\date{\today}

\begin{abstract}
Experiments on a thin layer of cohesive wet granular matter under vertical vibrations reveal kink separated domains that collide with the container at different phases. Due to the strong cohesion arising from the formation of liquid bridges between adjacent particles, the domains move collectively upon vibrations. Depending on the periodicity of this collective motion, the kink fronts may propagate, couple with each other and form rotating spiral patterns in the case of period tripling, or stay as standing wave patterns in the case of period doubling. Moreover, both patterns may coexist with granular `gas bubbles' -- phase separation into a liquidlike and a gaslike state. Stability diagrams for the instabilities measured with various granular layer mass $m$ and container height $H$ are presented. The onsets for both types of patterns and their dependency on $m$ and $H$ can be quantitatively captured with a model considering the granular layer as a single particle colliding completely inelastically with the container. 
\end{abstract}


\maketitle

\section{Introduction}

From Chladni figures~\cite{Chladni1787} to Faraday heaping~\cite{Faraday1831}, from dune formation in nature~\cite{Bagnold1941} to the segregation of granular mixtures in industries~\cite{Ottino1997}, pattern formation in granular matter has been attracting interest from physicists, engineers and geologists over centuries~\cite{Ristow2000,Aranson2006,Aranson2009}. In the past decades, many intriguing instabilities have been discovered, such as triggered avalanches~\cite{Daerr1999}, propagating fronts~\cite{Losert1999}, heap corrugation and transport~\cite{Falcon1999,Miao2006}, convection~\cite{Aoki1996,Liffman1997,Miao2004,Rietz2008,Fortini2015}, fingering in air~\cite{Pouliquen1997} or in water~\cite{Voeltz2001}, and stratification patterns of granular mixtures driven by avalanches~\cite{Gray1997} or by horizontal vibrations~\cite{Mullin2000,Reis2002}. Particularly, standing wave patterns in a vertically agitated dry granular medium have been investigated extensively, both in quasi-two-dimensional and three-dimensional systems~\cite{Clement1996,Melo1994}. Depending on the vibration strength, the system undergoes a period doubling cascade accompanied by stripe, square, hexagonal and spiral patterns as well as kink waves~\cite{Melo1995,Bizon1998,Aranson1999,Bruyn01,Zhang2005}. The dispersion relation of the subharmonic patterns was found to be reminiscent of gravity waves in a fluid~\cite{Melo1994}, suggesting the possibility to describe vibrofluidized granular matter as a continuum. Localized excitations, also coined as oscillons, were also  discovered in the hysteresis region of the primary square or stripe pattern forming instabilities~\cite{Umbanhowar1996}. 

Due to the strong cohesion arising from the formation of liquid bridges between adjacent particles, partially wet granular matter exhibits a different pattern forming scenario: Three armed rotating spiral pattern, which manifests a peculiar period tripling bifurcation, was found to dominate~\cite{Huang2011}. The spiral arms correspond to the kinks separating domains that collide collectively with the container at different phases. Further characterizations on the dynamics of spiral arms revealed that the rotation frequency is finite at the threshold and grows linearly with the peak vibrating acceleration~\cite{Huang2013}.  

As the granular layer moves collectively within the container, any other periodicity beside 3 may well exist in the system. Thus, it is intuitive to seek for patterns with various periodicities so as to find a clue on what determines the periodicity and why the period tripling state is preferred. Here we present the stability diagrams measured with a systematic variation of two associated parameters: Granular layer mass $m$ and container height $H$. We find, in addition to period tripling patterns, period doubling standing wave patterns as well as phase separations~\cite{Fingerle2008,Huang2009}. Both period doubling and tripling regions found in the experiments match the predictions of a single particle bouncing model considering collisions with both the bottom and the lid of the container, illustrating the roles that the two parameters play in determining the periodicity. Thus, this investigation provides possible pathways to control the periodicity of patterns in vertically agitated wet granular matter.

\section{Experimental Set-up}
\label{sec:meth}

\begin{figure}
\includegraphics[width = 0.65\columnwidth]{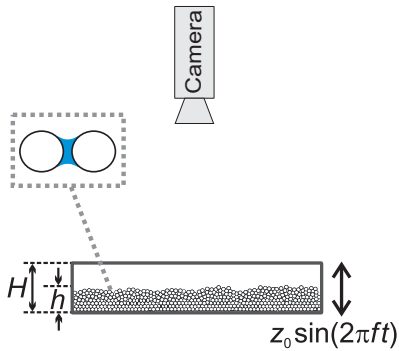}
\caption{\label{setup}(Color online) Schematic of the experimental set-up with definitions of the granular layer thickness $h$ and the container height $H$. The magnified view of the wet granular sample illustrates a liquid bridge formed between two adjacent particles. $z_{0}$, $f$, and $t$ denote maximum vibration amplitude, frequency, and time in the laboratory frame correspondingly. }
\end{figure}

Figure~\ref{setup} shows a schematic of the experimental set-up. Cleaned glass beads (SiLiBeads S) with a diameter of $d=0.78$\,mm and 10~\% polydispersity, after being mixed with purified water (specific resistance $18.2$\,M$\Omega\cdot$cm, \mbox{LaborStar TWF}), are added into a cylindrical container with a fixed mass of $768$\,g, an inner radius $R=8$\,cm and a height $H$. The filling fraction of the particles is defined as $\mathit{\Phi}=m/(\pi R^2 \rho_{\rm g} H)$, where $\rho_{\rm g}=2.50\,{\rm g}\cdot{\rm cm}^{-3}$ is the density of the glass beads. The height of the granular sample is estimated with $h\approx m/(\pi R^2 \rho_{\rm g} \eta)$, where $\eta$ corresponds to the packing density. Here the density of random close packing $\eta=0.64$ is used as an approximation. The sample is kept within the pendular state (i.e., cohesion arising mainly from liquid bridges formed between adjacent particles at contact) through keeping the liquid content $W=V_{\rm w}/V_{\rm g}\approx1.6$~\%, where $V_{\rm w}$ and $V_{\rm g}$ are the volume of the wetting liquid and that of the glass beads correspondingly. The container is agitated vertically against gravity with an electromagnetic shaker (Tira TV50350). The frequency $f$ and amplitude $z_{0}$ of the sinusoidal vibrations are controlled with a function generator (Agilent FG33220) and the dimensionless acceleration $\mathit{\Gamma}=4\pi^2 f^2 z_{0}/g$ is measured with an accelerometer (Dytran 3035B2), where $g$ is the gravitational acceleration. The collective behavior of the sample is captured with a high speed camera (IDT MotionScope M3) mounted above the container. The camera is triggered by a synchronized multi-pulse generator to capture images at fixed phases of each vibration cycle. More details on the set-up control can be found in Ref.~\cite{Huang2011}.

\section{Stability diagram}

\begin{figure}
\includegraphics[width = 0.95\columnwidth]{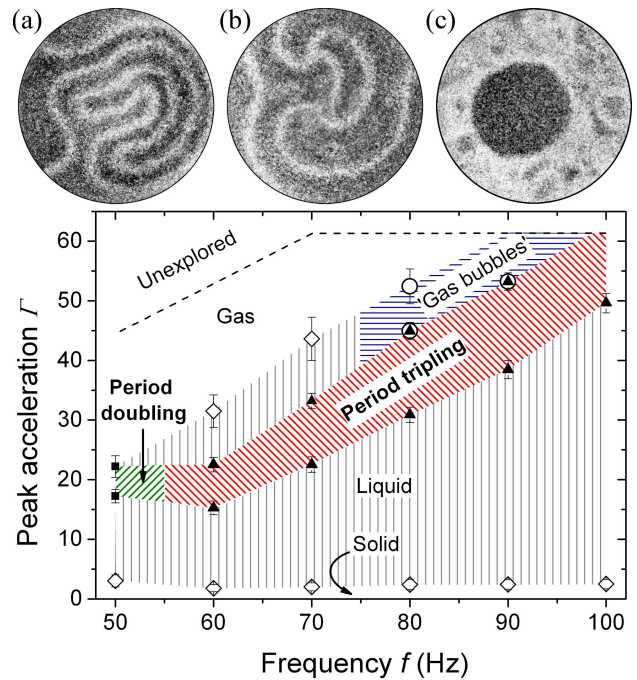}
\caption{\label{pd}(Color online) Stability diagram showing the collective behavior of the granular sample in the experiments. (a), (b) and (c) are typically snapshots for period doubling standing wave pattern, period tripling rotating spirals, and granular `gas bubbles' (i.e., the coexistence of a liquid- and a gaslike state). The rotating spiral pattern shown in (b) is an average of three images taken at consecutive vibration cycles to enhance the contrast. Other parameters: $m=113$\,g, $h\approx3.5$\,mm, $H=10.6$\,mm, and $\mathit{\Phi}=0.21$.}  
\end{figure}

\begin{figure*}
\includegraphics[width = 0.85\textwidth]{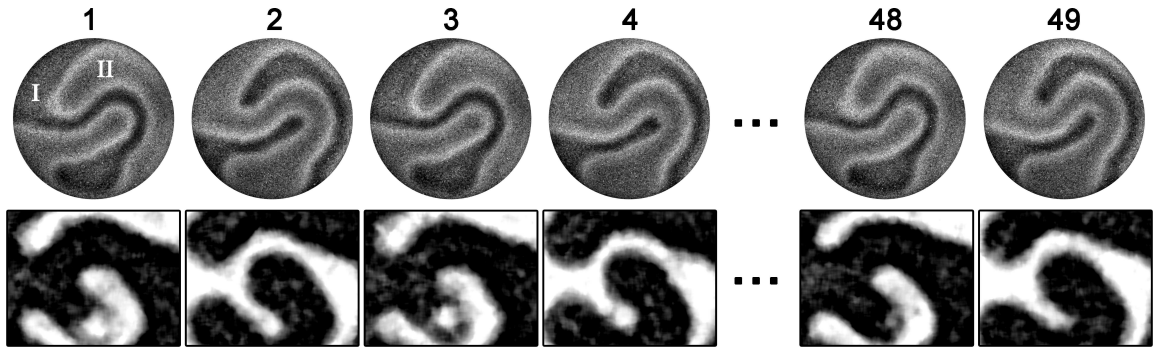}
\caption{\label{sa} Upper panel: Snapshots of a typical period doubling pattern taken at a fixed phase of consecutive vibration cycles (indicated as numbers) with $f=50$\,Hz and $\mathit{\Gamma}\approx17.3$\,g. Lower panel: Spatially resolved covariance of subsequent images captured as the granular layer collides with the container in another run of experiment with a closer view on the center region of the container. The brightness is related to the averaged mobility of the particles.}
\end{figure*}

\begin{figure}
\includegraphics[width = 0.75\columnwidth]{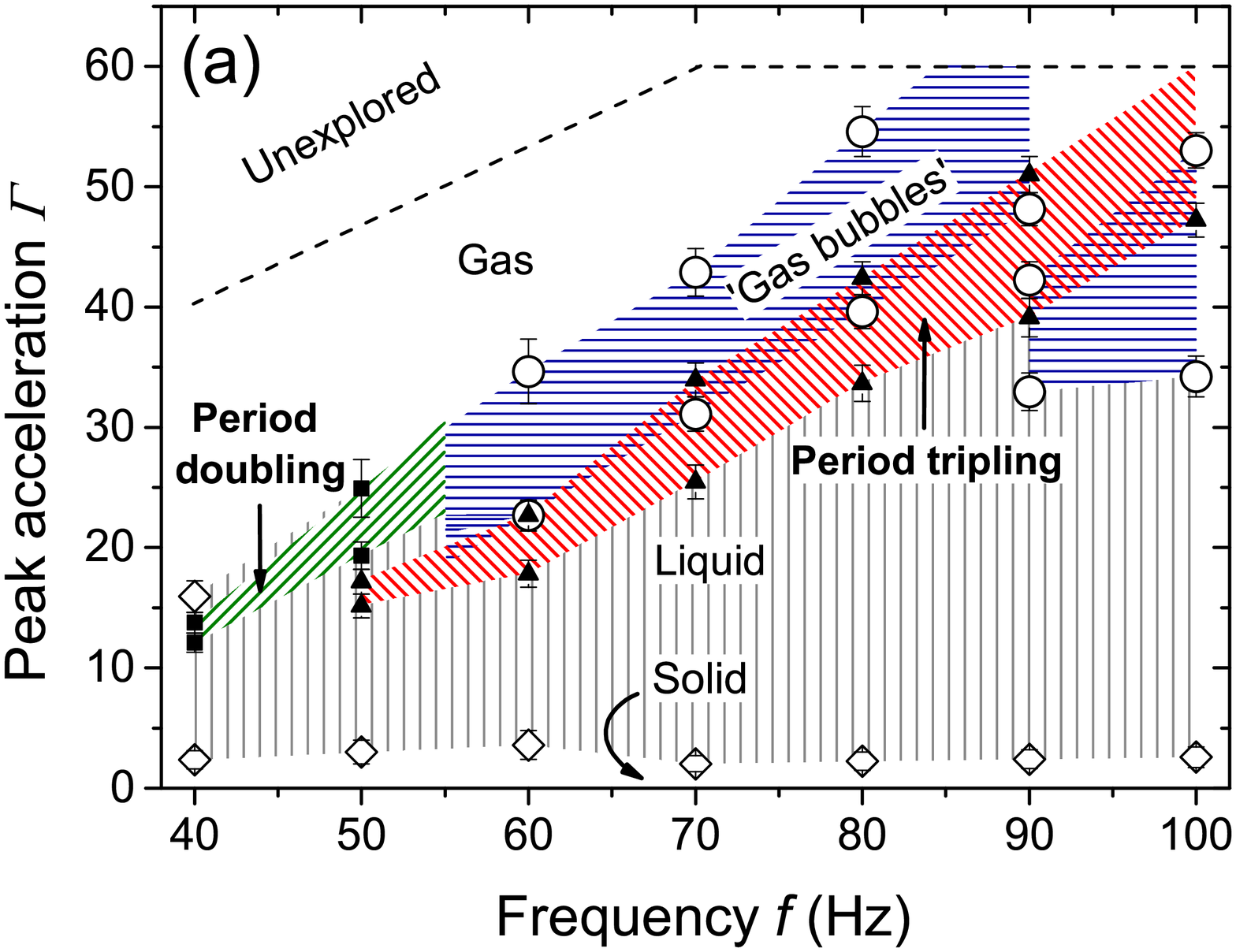} \\
\vskip 1ex
\includegraphics[width = 0.75\columnwidth]{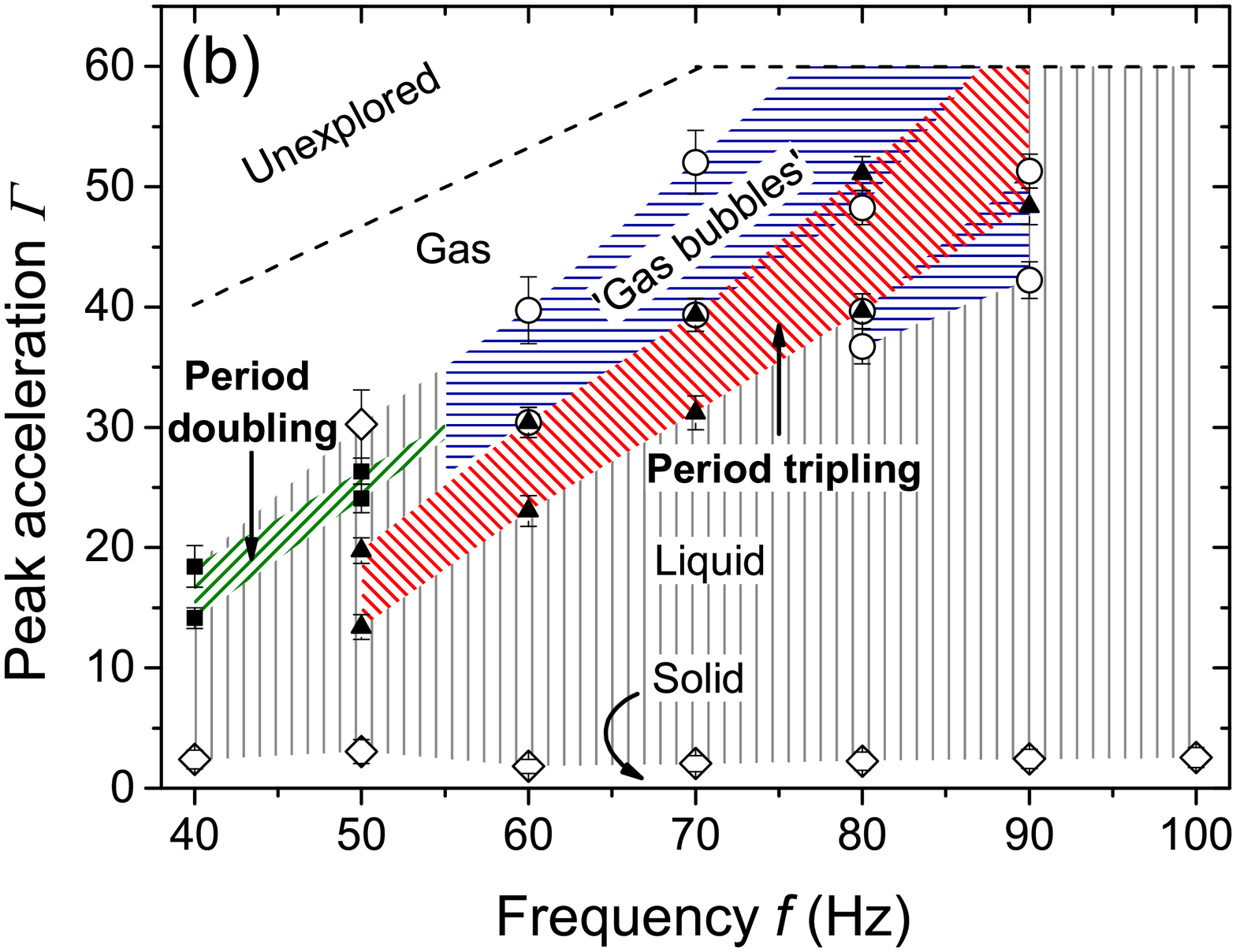} \\
\vskip 1ex
\includegraphics[width = 0.75\columnwidth]{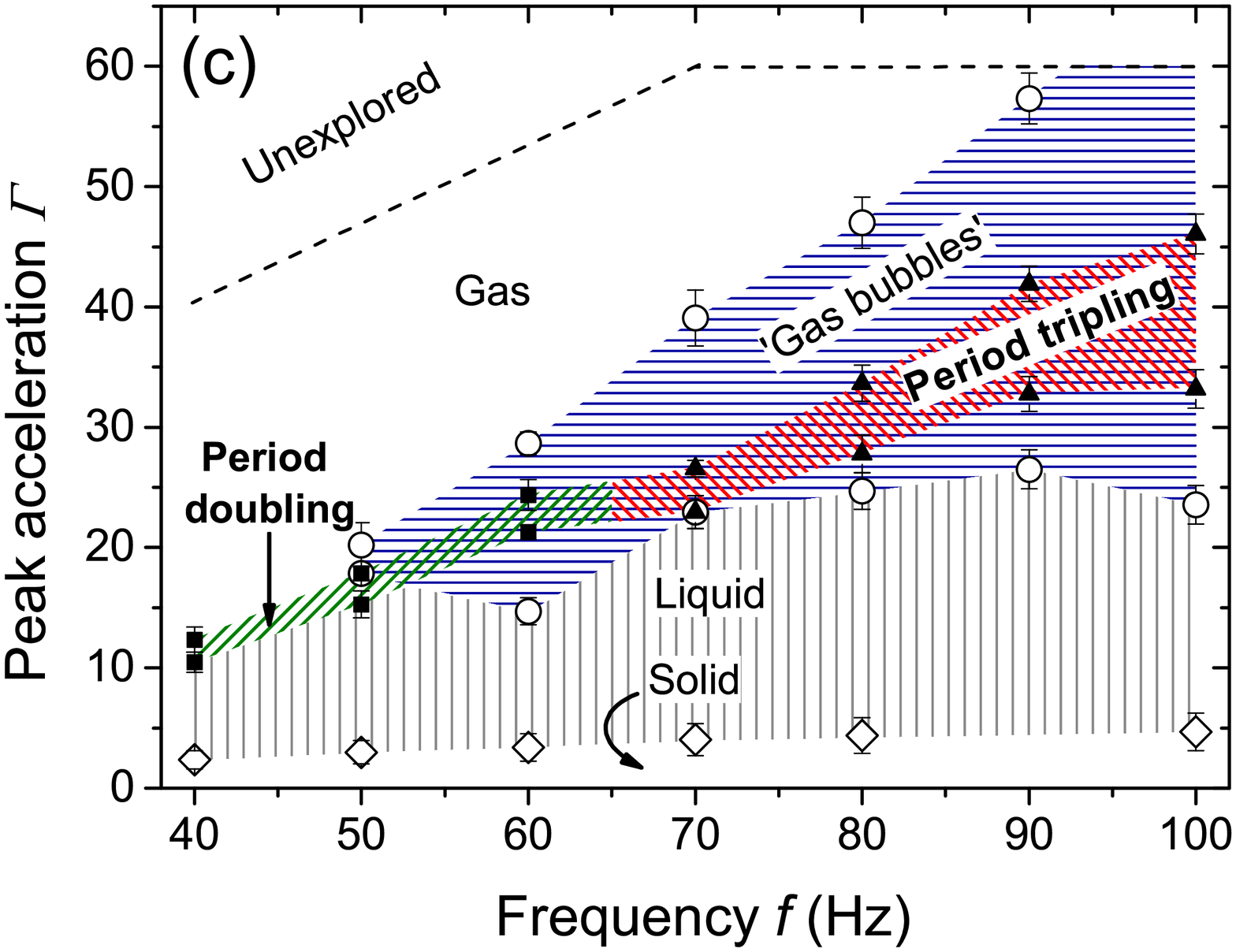} 
\caption{\label{pdcmp}(Color online) Stability diagrams for various combinations of sample mass $m$ and container height $H$ but the same filling fraction $\mathit{\Phi}=0.18$: (a) $m=96$\,g, $h\approx3.0$\,mm and $H=10.6$\,mm; (b) $m=111$\,g, $h\approx3.5$\,mm and $H=12.3$\,mm; (c) $m=81.6$\,g, $h\approx2.5$\,mm and $H=9.0$\,mm. The boundaries between various regimes are obtained the same way as in Fig.~\ref{pd}. Overlapping between various regimes corresponds to coexistence.}
\end{figure}

Figure~\ref{pd} shows the stability diagram measured with decreasing $\mathit{\Gamma}$ at various $f$. Increasing $\mathit{\Gamma}$ yields quantitatively the same threshold for the onset of patterns. The boundaries between various states are determined through an average of three runs of experiments and the uncertainty corresponds to the step of $\mathit{\Gamma}$. There exists slight hysteresis ($\sim 10$~\%) for the critical acceleration at which the transition from a solidlike (i.e., no particles moving) to a liquidlike (i.e., particles moving around while keeping contacts with their neighbors) state, as well as the phase separation into a coexistence between a liquidlike and a gaslike state (`gas bubbles'), arises. Such a hysteretic behavior is in agreement with former investigations~\cite{Fingerle2008,Huang2009}. In the gaslike state, most of the particles move individually without contacts to the others, and the granular layer expands to fill the whole container. The melting transition is measured through monitoring the changes of subsequent top view images, as the transition tends to start from the free surface of the granular layer~\cite{Roeller2011,May2013}. 

If the vibration strength increases further, a standing wave pattern emerges at a relatively low vibration frequency $f=50$\,Hz. As the typical snapshot (a) illustrates, the pattern is composed of domains split by curved interfaces that appear bright. A measurement of the surface profile with the laser profilometry method~\cite{Raton95s} reveals that the bright interfaces correspond to the kinks separating regions with different heights. At even lower frequencies, visual inspection is hindered by particles sticking on the lid of the container. In contrast to the non-propagating kinks (i.e., standing waves), a rotating spiral pattern is found to dominate for $f\ge 60$\,Hz. As shown in (b), the typical pattern is composed of three armed rotating spirals or propagating fronts. The spirals typically have meandering cores and may couple with each other through sharing arms. Different from the standing wave pattern described above, the kink waves propagate as period tripling breaks the spatiotemporal symmetry~\cite{Huang2011}. 

As $\mathit{\Gamma}$ increases further above the pattern forming region, the system may evolve into a featureless liquidlike state or phase separate before entering the homogeneous gaslike state, depending on the vibration frequency. As shown in (c), the density in the gaslike region is much lower than the surrounding liquidlike region. Temporal fluctuations may occur in the liquidlike region, while no patterns are observed in the `gas bubble' region. The transition from the liquidlike to the gaslike state is determined from a sudden increase of the noise level of the signal from the accelerometer, because individual collisions between the particles and the container are much more prominent in the gaslike state in comparison to the liquidlike one. Such an enhanced noise level also leads to larger error for the onset of the gaslike state in comparison to the other state boundaries.

Figure~\ref{sa} shows the time evolution of a standing wave pattern. It is a labyrinth type pattern [see also Fig.~\ref{pd}\,(a)] and different runs of experiments yield dramatically different morphologies of the pattern. Here, it behaves like two fingers slightly wrapped around each other. The images in the upper panel taken at a fixed phase of subsequent vibration cycles indicate that the pattern has a periodicity that doubles that of the vibrating plate. This is similar to the period doubling patterns observed in agitated dry granular matter~\cite{Melo1995}. However, there are no periodic structures (e.g., squares or stripes) observed here. This difference might be attributed to the strong cohesion from the presence of liquid bridges: The injected momentum in the vibrating direction is more difficult to be transferred to the horizontal direction in comparison to the dry case. Moreover, the standing wave pattern is also reminiscent to the `phase bubbles' observed in vertically vibrated dry granular layers ~\cite{Moon2001} as they are both composed of kink separated domains vibrating with a phase difference. In comparison to the transient `phase bubbles' that shrink and disappear within tens to hundreds of vibration cycles after nucleation, the standing wave pattern observed here is more stable. Within a time scale of a few hundred vibration cycles, the morphology of the standing wave pattern stays the same. As time evolves up to $10^5$ vibration cycles, a target pattern sometimes arises.

A combination of the snapshots taken after the granular layer collides with the container (upper panel of Fig.~\ref{sa}) and spatially resolved covariance $C(x,y)$ (lower panel) reveals the dynamics of the pattern. $C(x,y)$ is calculated with

\begin{equation}
C(x,y)=\sum_{\Delta x, \Delta y=0}^{d}I_t(x+\Delta x,y+\Delta y) I_{t+\Delta t}(x+\Delta x,y+\Delta y),
\end{equation}

\noindent where $I_t(x,y)$ corresponds to the intensity of an image captured at time $t$, $\Delta t$ is the time step between consecutive frames, and $d$ denotes the particle diameter in pixels. In order to resolve the mobility of the particles, the images are captured with $10$ frames per vibration cycle [i.e., $\Delta t=1/(10f)$]. 

In the first vibration cycle, one side of the front (marked with I) collides with the container and collects kinetic energy there. Consequently, the mobility of particles in this region enhances so that an expansion into the neighboring region II is expected. In the next vibration cycle, the front will be pushed back as region II collides with the container. Therefore, the fronts swing back and forth at the time scale of vibration cycles. Such a process is better illustrated with $C(x,y)$ shown in the lower panel of Fig.~\ref{sa}, as the black (more mobile) and white (less mobile) regions switch at each vibration cycle. For the case of non-cohesive dry granular matter, the friction between the kink separated granular layers moving out of phase can induce convection in the vicinity of the kink region for both two-~\cite{Zhang2005} and three-dimensional~\cite{Moon2003} systems. For the case of wet granular matter with strong cohesion between adjacent particles, it is still unclear whether convection can be triggered or not. This will be a topic for further investigations.

As the granular layer collides with the container collectively, we speculate that the mass $m$ of the granular layer (or corresponding layer thickness $h$) and the height of the container $H$ play essential roles in determining the period for the granular layer to travel within the container. In order to explore such an influence, we compare stability diagrams measured with a systematic variation of both parameters.

Figure~\ref{pdcmp}\,(a) shows the stability diagram for the same container height $H=10.6$\,mm but smaller $m$ in comparison to Fig.~\ref{pd}. Such a change of the sample mass leads to the following features in the stability diagram: (i)~The period tripling pattern region expands to $f=50$\,Hz in a $\mathit{\Gamma}$ region smaller than that of the standing wave pattern. This suggests that the periodicity of the pattern is not solely determined by the vibration frequency. (ii)~The phase separation regime expands to both lower frequency ($f=60$\,Hz) and lower $\mathit{\Gamma}$, particularly in the high frequency regime. The enhanced probability for `gas bubbles' to nucleate can be attributed to the granular layer thickness: The thinner the granular layer, the lower the energy barrier for a `gas bubble' to nucleate because the number of liquid bridges to break along the vibration direction is reduced. (iii)~The period tripling pattern and `gas bubble' regimes can overlap, leading to a state in which propagating kink fronts or rotating spirals are observed in the liquidlike region surrounding the `gas bubbles'.

Moreover, we discuss the influence of $H$ through a comparison between Fig.~\ref{pdcmp}\,(b) and Fig.~\ref{pd}, as they share roughly the same $m$. As $H$ increases, the phase separation region expands and the thresholds for both patterns grow faster with $f$. The latter feature suggests that the influence of $H$ might be associated with the departure velocity ($\sim\mathit{\Gamma}/f$) required for the granular layer to reach the container lid. The strong influence of $H$ indicates that a collision between the granular layer and the container lid is essential for the observed patterns. Note that the maximum elevation height of a particle bouncing completely inelastically with the container (see Sec.~\ref{sec:model} for details) is typically much larger than the container height in the pattern forming region.

The stability diagrams presented in Fig.~\ref{pdcmp} have the same filling fraction $\Phi$ but various $m$. A comparison of (b) or (c) with (a) reveals the `gas bubble' region shrinks (expands) as granular layer is thicker (thinner). This is similar to the above comparison between various $m$ but the same $H$. For a granular layer thickness of roughly $3$--$4$ particle diameters (c), the `gas bubble' region expands further into the period doubling region. For $f\ge50$\,Hz, all pattern forming instabilities coexist with phase separation; that is, they appear in the liquidlike phase surrounding the `gas bubbles'. Moreover, the comparison also reveals that the onset for both patterns grows faster (slower) with $f$ as $m$ increases (decreases), reminiscent of the above comparison between various $H$ but the same $m$. The similarities to the variation of $m$ or $H$ individually suggest that the influence of $m$ and $H$ is not coupled with each other through the filling fraction.

\section{Model}
\label{sec:model}

\begin{figure}
\includegraphics[width = 0.75\columnwidth]{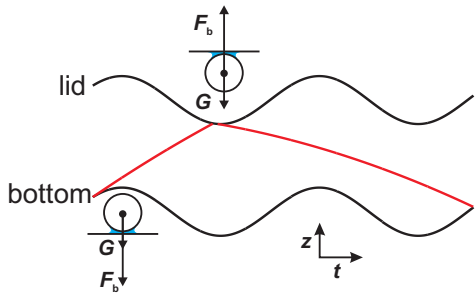}
\caption{\label{model}(Color online) A sketch of the model considering the granular layer as a single particle undergoing completely inelastic collisions with the lid and bottom of the container. The red (gray) trajectories correspond to the free flying period of the particle. $\vec{G}$ and $\vec{F_{\rm b}}$ denote the weight of the granular layer and the cohesive force between the granular layer and the container.}
\end{figure}

As shown in Fig.~\ref{model}, the model considers the granular layer as a single wet particle colliding completely inelastically with the container. The weight of the particle is $\vec G = - m g \vec e_{\rm z}$ with $\vec e_{\rm z}$ a unit vector pointing in the $z$ direction. The magnitude of the cohesive force is estimated with $F_{\rm b}=\pi N_{\rm c} \sigma d \cos{\theta}$, where $N_{\rm c}=4 \eta_{\rm 2D} R^2/d^2$ is the number of contacts between the granular layer and the container, $\sigma=0.072$\,N/m is the surface tension of water, and $\theta=0$ is the contact angle. Here the area fraction $\eta_{\rm 2D}=0.84$ is estimated with the random close packing of spheres in two dimensions~\cite{Desmond2009}. The total force acting on the particle gives rise to two critical accelerations: $a_{\rm l}=(n-1)g$ for the container lid and $a_{\rm b}=-(n+1)g$ for the container bottom, where $n=F_{\rm b}/(mg)$ is the ratio between the cohesive force and the gravitational force acting on the particle. 

In the numerical analysis, the particle is placed initially on the container bottom. As the criterion for the acceleration of the container $a=-4\pi^2 f^2 z_{0} \sin{2\pi f t}\le a_{\rm b}$ is fulfilled, the particle starts to detach from the container bottom. If the detachment eventually leads to a break of the capillary bridge, the rupture process succeeds and the particle starts a parabolic flight. The trajectory of the particle during the rupture process is calculated considering the decay of the capillary force with increasing separation distance~\cite{Willett2000}. From the trajectory of the parabolic flight and that of the container bottom or the lid, we determine the next collision time numerically with Brent's algorithm~\cite{Brent1973}. Considering the thickness of the granular layer, the container height is chosen to be $H-h$ while estimating the collision time with the lid. After colliding with the container, the particle may start to detach immediately from or stick to the container, depending on whether the criterion $a\ge a_{\rm l}$ for the lid or $a\le a_{\rm b}$ for the bottom is fulfilled or not. If not, the particle moves together with the container until the criterion is satisfied. Subsequently the rupture process starts and the particle continues to bounce in the container. Based on the number of vibration periods the particle takes until it sticks back to the container bottom, we estimate the preferred periodicity for the wet granular layer under a certain vibration frequency and peak acceleration. 

\begin{figure}
\includegraphics[width = 0.85\columnwidth]{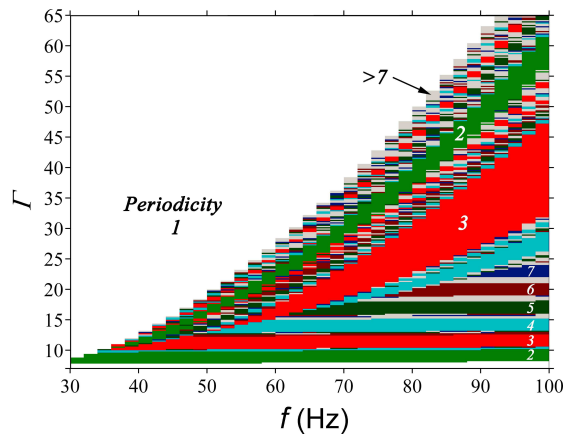}
\caption{\label{pdmodel}(Color online) Periodicity predicted by the numerical analysis based on the single particle bouncing model. The parameters are chosen to match those in Fig.~\ref{pd}.}
\end{figure}

As shown in Fig.~\ref{pdmodel}, the numerical analysis predicts two classes of periodicity regions: One stays constant with $f$ and the periodicity increases step by step with $\mathit{\Gamma}$, which we call periodicity class I. The other one, called periodicity class II, starts at a certain frequency $f_{\rm c}$ and grows with $f$ monotonically, and the periodicity decreases as $\mathit{\Gamma}$ grows. The spaces between neighboring periodicity regions are chaotic. This is manifested by the large absolute values as well as fluctuations of the periodicity arising from the immediate detachment of the particle after colliding with the container. 

Qualitatively, the class II periodicity regions agree with the pattern forming regions in the stability diagrams shown in Fig.~\ref{pd} and~\ref{pdcmp} on the following features: (i)~There exists a certain $f_{\rm c}$ above which a certain periodicity region starts. (ii)~Both boundaries of a certain periodicity region grow with $f$. (iii)~The period tripling region dominates and expands as $f$ grows. (iv)~The period doubling region starts at a lower $f_{\rm c}$ and grows in a $\mathit{\Gamma}$ region larger than the period tripling one. Different from the experiments, the model predicts an additional region with periodicity four below the period tripling one.

In order to have a more quantitative understanding of the periodicity diagram and its dependency on the control parameters, we have varied both $m$ and $H$ the same way as the experiments. As a starting point, we explain why the periodicity is independent of $f$ in periodicity class I. Following the above description, the granular layer starts to detach from the container bottom as the acceleration of the plate satisfies $a\le a_{\rm b}=-(n+1)g$ in the laboratory frame. Assuming that the free flight period starts immediately (i.e., neglecting the rupture process), we can estimate the velocity at departure with

\begin{figure*}
\includegraphics[width = 0.4\textwidth]{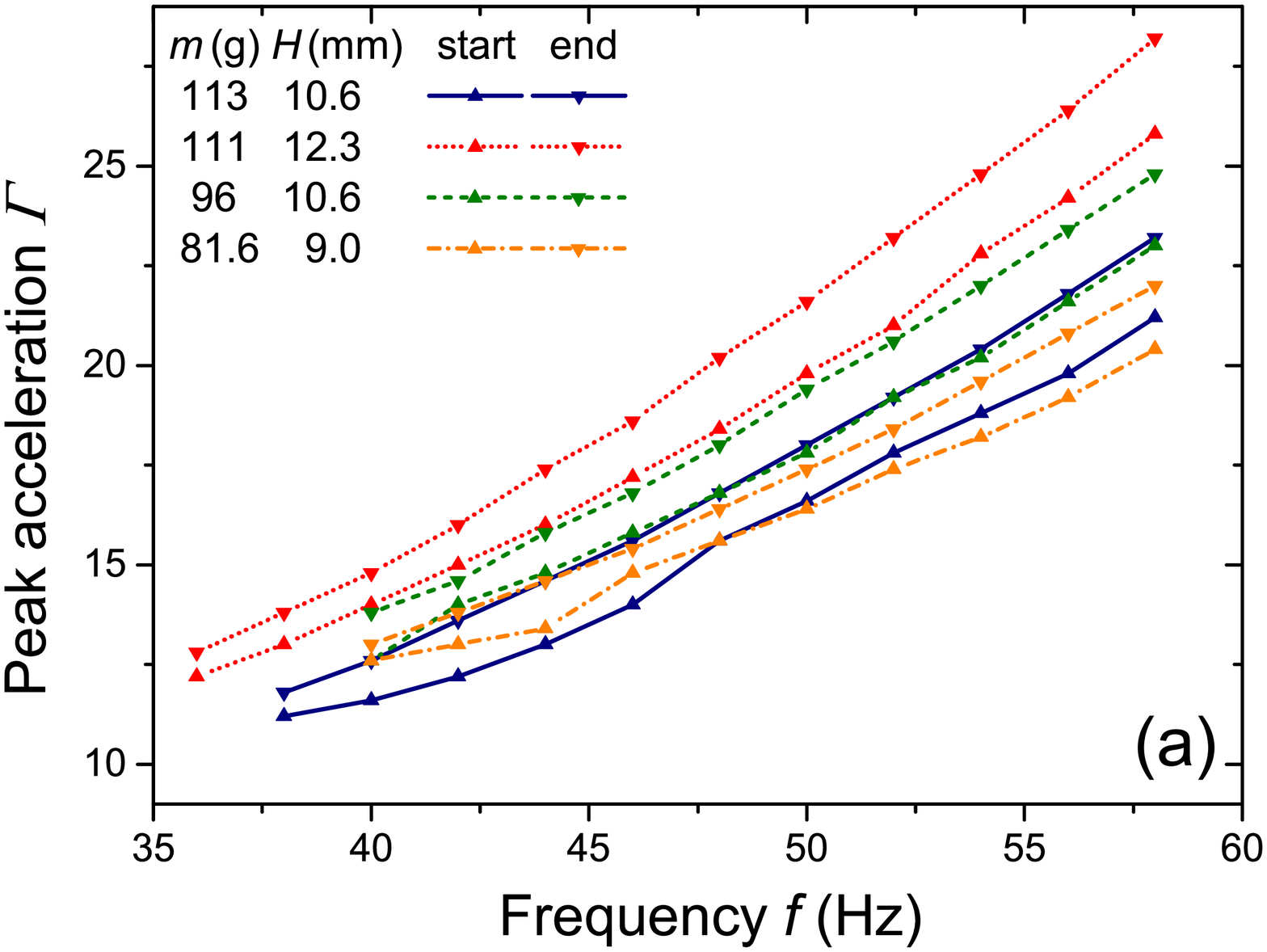} \hskip 1em
\includegraphics[width = 0.4\textwidth]{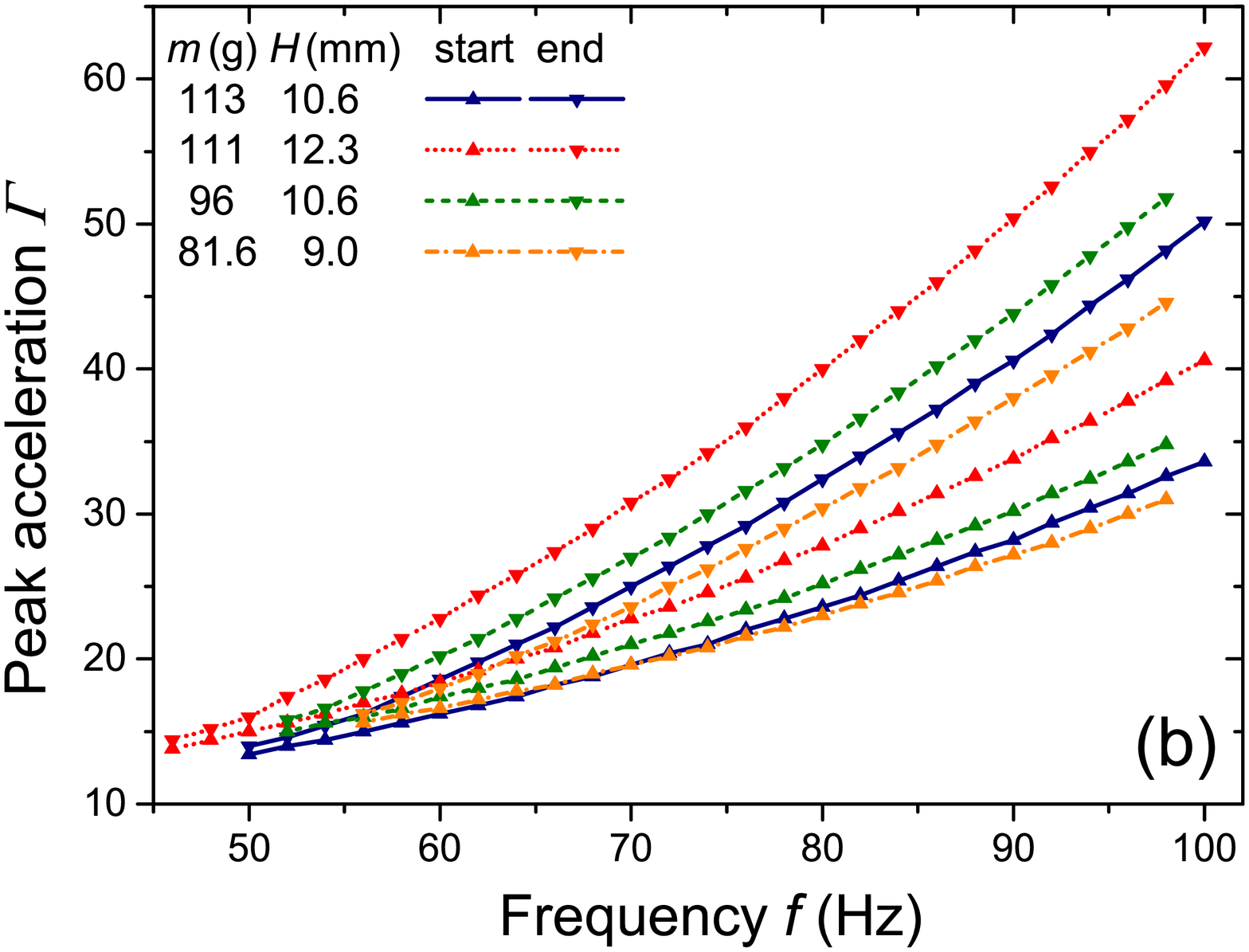} \\
\vskip 2em
\includegraphics[width = 0.4\textwidth]{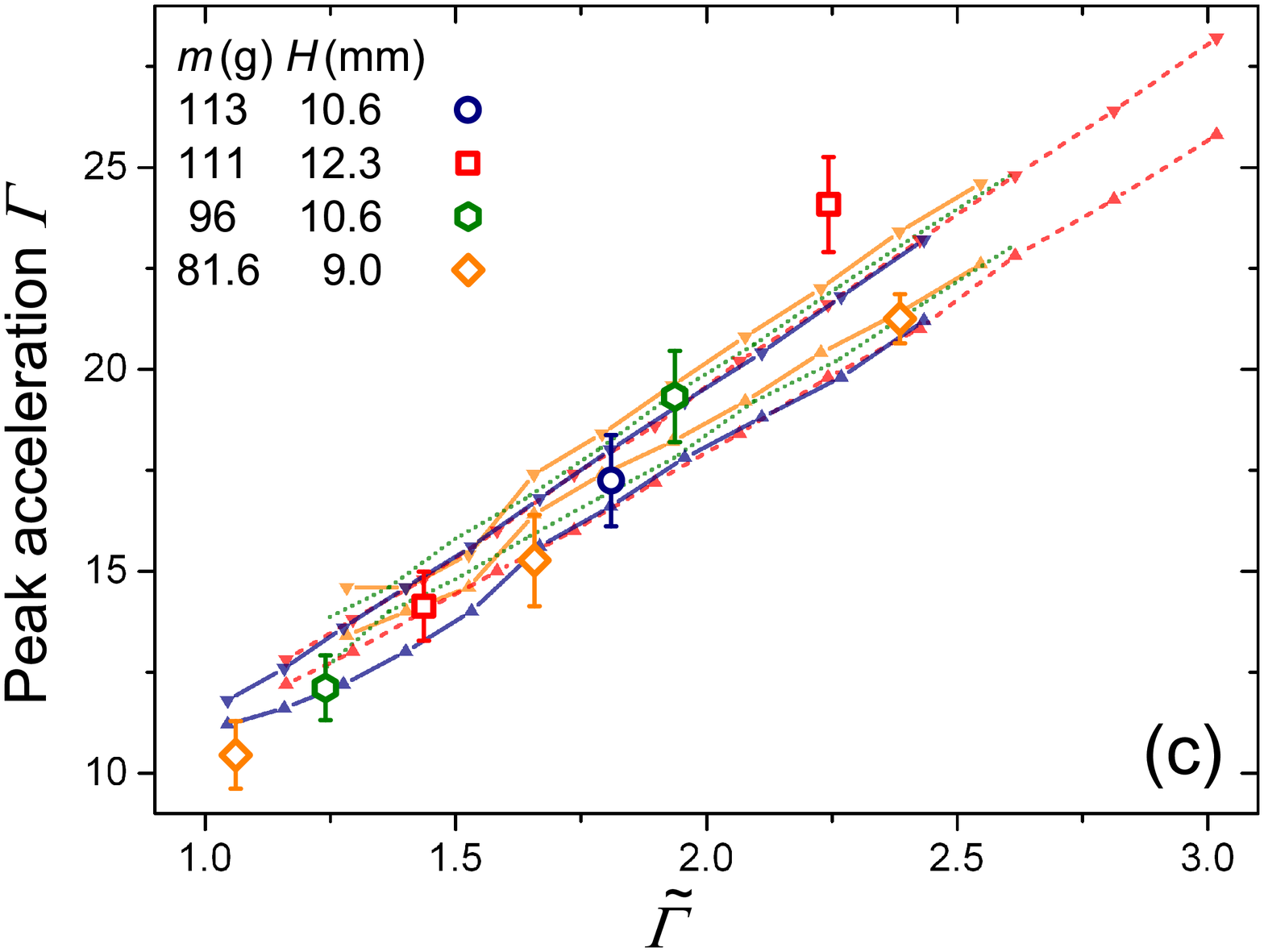} \hskip 1em
\includegraphics[width = 0.4\textwidth]{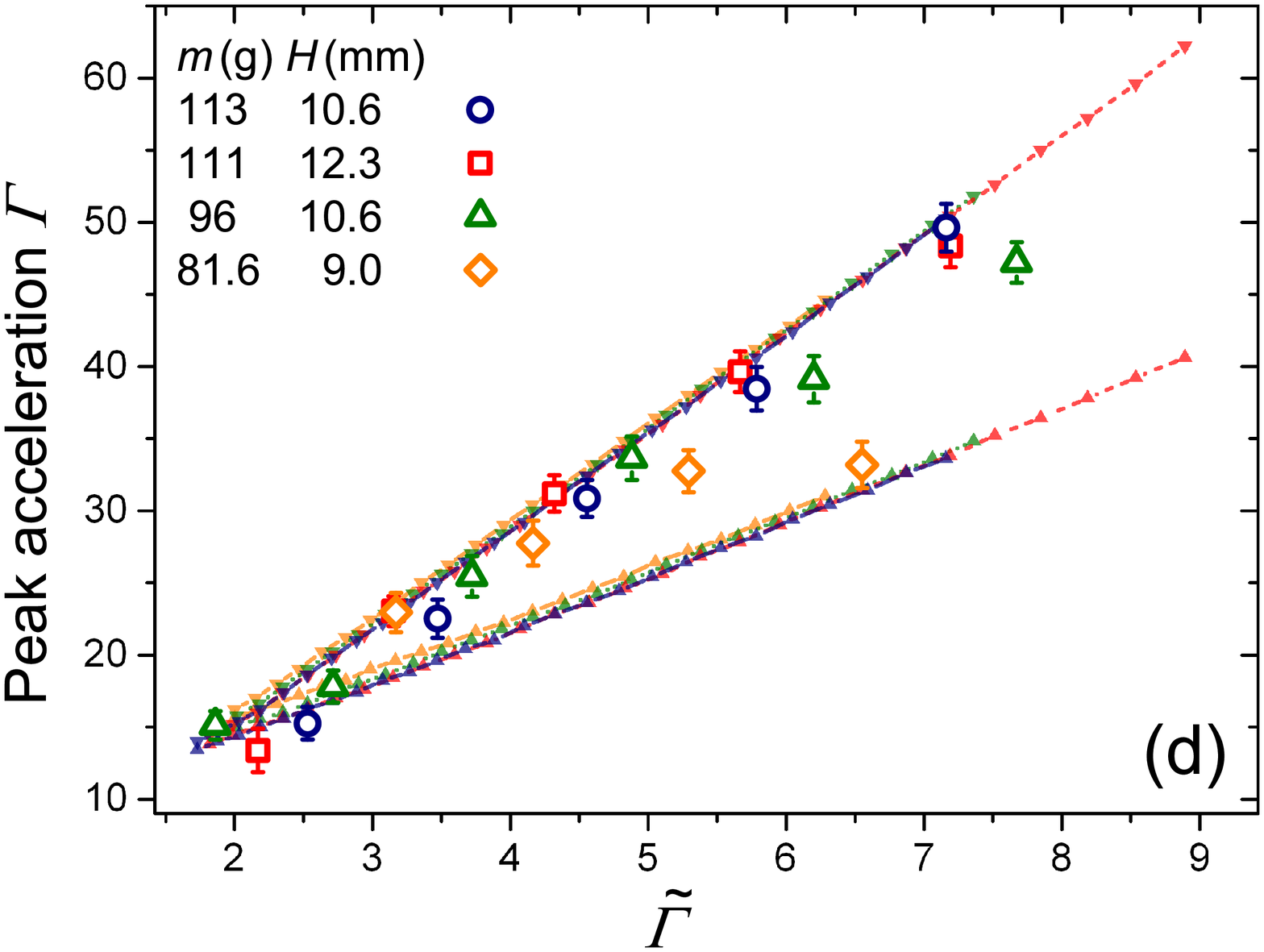} 
\caption{\label{scaling}(Color online) Boundaries for period doubling (a) and tripling (b) regions from the numerical analysis. Various types of connection lines correspond to various combinations of parameters $m$ and $H$, which are the same as in the experiments. Comparisons between the predictions of the numerical analysis [connected symbols, same legend as in panels (a) and (b)] and the onsets for both period doubling (c) and tripling (d) patterns measured experimentally in the {$\mathit{\Gamma}$--$\mathit{\tilde\Gamma}$} plane. See the text for a definition of the dimensionless control parameter $\mathit{\tilde\Gamma}$.}
\end{figure*}

\begin{equation}    
v_{\rm c}=\sqrt{(2\pi f z_{\rm 0})^2-\left(\frac{a_{\rm b}}{2\pi f} \right)^2},
\end{equation}

\noindent where $z_{\rm 0}$ corresponds to the peak vibration amplitude. Using the definition of $\mathit{\Gamma}$ and $a_{\rm b}$, we have

\begin{equation}    
\label{eq:vc}
v_{\rm c}=\frac{g}{2\pi f}\sqrt{\mathit{\Gamma}^2-(n+1)^2},
\end{equation}
 
\noindent which shows its dependency on the control parameters $f$, $\mathit{\Gamma}$ and $n$. Note that the dependence on $n$ represents the influence from sample mass $m$. As a first approximation, we further ignore the influence of vibration amplitude on the free flying period of the particle [i.e., assuming $z_{\rm 0} \ll (H-h)$]. Consequently, the periodicity of the parabolic flight $T$ can be estimated with

\begin{equation}    
\label{eq:per}
T=\lceil\frac{2v_{\rm c}}{g}\cdot f \rceil=\lceil\frac{1}{\pi}\sqrt{\mathit{\Gamma}^2-(n+1)^2}\rceil,
\end{equation}
 
\noindent which is independent of $f$, and grows with $\mathit{\Gamma}$ as well as $n$. If we take the parameters used in Fig.~\ref{pdmodel} and skip the ceiling function $\lceil \rceil$, Eq.~\ref{eq:per} leads to $\mathit{\Gamma}\approx9.1$ and $11.5$ for periodicity $T=2$ and $3$, which fall into the middle of the ranges predicted by the numerical analysis (8.2--10.2 for periodicity 2 and 10.8--12.8 for periodicity 3). Thus, the periodicity class I corresponds to the situation without collisions with the lid of the container. 

As the above analysis applies as well to the cohesionless case with $n=0$, Eq.~\ref{eq:per} gives us a clue on why the onset of subharmonic patterns observed in dry granular matter~\cite{Melo95} is weakly dependent on the vibration frequency. However, no instabilities are observed in this class of periodicity for the cohesive case that we focus on here, presumably due to the strong cohesion between adjacent particles as well as between the granular layer and the container. The granular layer may follow the periodicity collectively, but the energy injection is not sufficient to effectively mobilize the particles and to initiate instabilities.

Following a similar approach, we can understand how the lid plays a role in determining the periodicity class II. Taking the same assumption of $z_{\rm 0}\ll(H-h)$, the criteria for the granular layer to reach the lid is $v_{\rm c}\ge \sqrt{2g(H-h)}$. Together with Eq.~\ref{eq:vc}, we obtain the \-corresponding 

\begin{equation}    
\label{eq:gamlid}
\mathit{\Gamma}=\sqrt{\frac{8\pi^2f^2(H-h)}{g}+(n+1)^2},
\end{equation}

\noindent at which the particle starts to reach the lid. From there on, we expect the periodicity to decrease with the growth of $\mathit{\Gamma}$, because the time for the particle to travel within the container reduces. Qualitatively, Eq.~\ref{eq:gamlid} explains the growth of the threshold acceleration $\mathit{\Gamma}$ with $f$ and $H$ observed in the experiments. Quantitatively, we cannot estimate the $\mathit{\Gamma}$-$f$ relation for various periodicities the same way as in the other case, because the periodicity relies not only on the free flying periods but also on the time that the particle travels together with the container.

Focusing on periodicity class II, we compare the periodicity regions predicted for various $m$ and $H$ by the numerical analysis in Fig~\ref{scaling}(a). It shows that either increasing $H$ or decreasing $m$ lead to enhanced onsets for both period doubling and tripling regions. This is in agreement with Eq.~\ref{eq:gamlid} (note that decreasing $m$ is equivalent to increasing $n$). 

Moreover, Eq.~\ref{eq:gamlid} suggests a new parameter $\mathit{\tilde\Gamma}=f^2(H-h)/g$ that combines the influence of both $f$ and $H-h$. It corresponds to a dimensionless acceleration where the amplitude is replaced with the maximum free traveling distance $H-h$ of the granular layer. As Fig.~\ref{scaling}(b) demonstrates, a replot of the numerical results shown in Fig.~\ref{scaling}(a) in the $\mathit{\Gamma}$--$\mathit{\tilde\Gamma}$ plane leads to a collapse of the data points for a certain periodicity. Similar data collapse is found for the onset of patterns measured experimentally, except for the case of $m=81.6$\,g at relatively large $\mathit{\tilde\Gamma}$. The data collapse demonstrates the coupling between various control parameters ($m$, $f$ and $H$) concerning the onsets of various periodicities in the system (note the relation between $h$ and $m$ shown above). 

A comparison between the numerical analysis and the experiments shows that the onset of period tripling patterns takes place at the upper bound of the periodicity region predicted by the numerics. This suggests that the periodicity is only a precondition for the patterns to appear. Following the above argument on why there are no patterns observed in periodicity class I regime, we speculate that sufficient energy injection is necessary to generate kinks, which are the basic ingredient for the standing waves as well as rotating spiral patterns observed experimentally.

Moreover, good agreements between the experimental and numerical results on the critical $\mathit{\tilde\Gamma}_{\rm c}$ above which a certain periodicity region emerges are found in Fig.~\ref{scaling}(b). $\mathit{\tilde\Gamma}_{\rm c}$ can be estimated quantitatively as follows. From Eq.~\ref{eq:per}, we estimate the acceleration to reach a certain periodicity $T$ at a fixed $n$

\begin{equation}    
\label{eq:gam1}
\mathit{\Gamma}\approx\sqrt{\pi^2 T^2+(n+1)^2},
\end{equation}

\noindent Supposing this is the acceleration at which the granular layer starts to touch the lid and the periodicity starts to decrease, we obtain, together with Eq.~\ref{eq:gamlid}, the critical frequency $f_{\rm c}$ at which the transition to periodicity $T-1$ starts:

\begin{equation}    
\label{eq:fc}
f_{\rm c}\approx T \sqrt{\frac{g}{8(H-h)}}.
\end{equation}

\noindent At the scale of the control parameter $\mathit{\tilde\Gamma}$, we estimate $\mathit{\tilde\Gamma}_{\rm c}$ for periodicity 2 or 3 to be $T^2/8=1.1$ or $2.0$, which agrees fairly well with the numerical ($1.0$ or $1.8$) and the experimental ($1.1$ or $1.9$) results shown in Fig.~\ref{scaling}(b).

\section{Conclusions}

In conclusion, the collective behavior of a thin layer of cohesive granular matter under vertical vibrations is investigated experimentally. Stability diagrams for standing wave patterns, rotating spirals and phase separations (granular `gas bubbles') are presented. For the period doubling induced standing wave patterns, the granular layer splits into two domains that collide with the container at different phases. The interface between the domains corresponds to kink fronts that swing back and forth with time, in contrast to the case of period tripling induced rotating spirals~\cite{Huang2011}. A comparison between the stability diagrams measured with various parameters reveals that the phase separation region expands as the sample mass $m$ (or correspondingly granular layer thickness $h$) decreases or the container height $H$ increases. Moreover, the influence of these parameters on various pattern forming regions is discussed through comparisons between various stability diagrams.

Using a simplified model considering the whole granular layer as a single particle colliding completely inelastically with the container, we predict various periodicity regions and compare with the stability diagram measured experimentally. For both standing wave and rotating spiral patterns, we find corresponding period doubling and tripling regions in the periodicity diagram predicted by the model. Further comparisons indicate that additional energy injection through colliding with the lid of the container is a precondition for both period doubling and tripling patterns to appear. Quantitatively, we introduce a dimensionless parameter $\mathit{\tilde\Gamma}$ that characterizes the dependency of the boundaries for various periodicity regions on the vibration frequency $f$, height of the granular layer $h$ as well as height of the container $H$. Moreover, the critical frequency $f_{\rm c}$ above which a certain periodicity region starts can also be rationalized with the model.

According to the model, the system could in principle generate patterns with a controlled number of periodicity (e.g., four-phase patterns~\cite{Elphick1998, Lin2000, Marts2004}) that may lead to novel pattern forming scenarios. Following the predictions from the model, a variation of the control parameters to explore possible period four instabilities will be a focus of further investigations. Moreover, we would also like to explore the possibility to control the dynamics of the domain interfaces with biharmonic driving~\cite{Aranson1999, Quintero2010}.

\begin{acknowledgments}  
The authors would like to thank Frank Gollwitzer for his preliminary work on the experimental set-up. Inspiring discussions with Ehud Meron and Alois W\"urger are gratefully acknowledged. This work is supported by the Deutsche Forschungsgemeinschaft through Grant No.~HU1939/2-1. 
\end{acknowledgments}

%

\end{document}